\documentclass[preprint,aps]{revtex4}

\usepackage{amsmath}

\usepackage{graphicx}

\begin{document}

\title{Thermodynamic potential of a mechanical constitutive model for two-phase band
flow}

\author{Katsuhiko Sato$^1$, Xue-Feng Yuan$^2$ and
Toshihiro Kawakatsu$^1$} \date{\today} \address{ $^1$ Department of Physics, Tohoku
University, Sendai 980-8578, Japan} \address{ $^2$ School of Chemical Engineering and
Analytical Science, The University of Manchester, Manchester, M60 1QD, United
Kingdom}

\begin{abstract} Starting from a simple mechanical constitutive model (the
non-local diffusive Johnson-Segalman model; DJS model), we provide a rigorous theoretical
explanation as to why a unique value of the stress plateau of a highly sheared
viscoelastic fluid is stably realized. The present analysis is based on a reduction theory
of the degrees of freedom of the model equation in the neighborhood of a critical
point, which leads to a time-evolution equation that is equivalent to those for
first-order phase transitions.  \end{abstract} 


\maketitle

Highly sheared viscoelastic fluids are known to have spatially inhomogeneous structures, i.e., ''shear bands'' \cite{kn:Hoffmann,kn:review-of-Cates}, where the fluid is separated into two regions with different values of the velocity gradient. In this phenomenon, a stress plateau
is observed in the stress-strain curve (SS curve).

This phenomenon has been explained using an N-shaped SS curve, as depicted in
Fig. 1 \cite{cates-mcleish}.  In the region of the shear rate where the SS curve has a
negative slope (referred to hereafter as the negative-slope region), the uniform
flow is unstable, and then the fluid separates into two stable domains having two
unique values of shear rates, denoted by $\dot{\gamma}_1$ and $\dot{\gamma}_2$
($\dot{\gamma}_1 < \dot{\gamma}_2$). {As the applied shear rate $\dot\gamma$
increases, the relative volume fraction of the domain with $\dot{\gamma}_2$
increases. This scenario enables us to explain the appearance of the shear stress
plateau using a consideration similar to thermodynamics. The validity of this
scenario has been generally confirmed by direct experimental
observations \cite{kn:birefringence,kn:NMR,kn:light-scattering}.}

On the other hand, there are some theoretical disadvantages associated with this scenario. One
such disadvantage is the fact that the origin of the unique value of the stress
plateau, $\sigma_s$, is unknown. (In experiments, the value of the stress plateau is
known to be uniquely determined independently of the flow history.)  A solution to
this problem is given by adding a non-local term (a diffusion term) to the
constitutive model. In References \cite{kn:yuan-Europhys,kn:Lu,kn:Fielding} the
uniqueness of the value of $\sigma_s$ is demonstrated numerically, and in
Reference \cite{Dhont} a selection rule for $\sigma_s$, which is analogous to the
Maxwell equal area construction, is provided analytically.

Although the problem of the unique determination of $\sigma_s$ was proven in the
above references, another important problem remains, i.e., the global stability
of the banded state. As stated above, in the negative-slope region the banded state
is finally realized because of the instability of the homogeneous flow. However, in
the remaining regions, (a)-(c) and (d)-(b) as indicated in Fig. 1, the SS curve has a
positive slope and so the homogeneous flow is stable (at least locally). Thus, it
is not guaranteed that the banded state is finally realized in these regions. In
other words, we cannot determine whether the homogeneous flow or the
banded flow is more stable.  This is because we have not yet had an evaluation function such as a
thermodynamic potential for the models describing the shear banding.

In the present article, we will demonstrate the metastability of the above-mentioned
homogeneous flow by deriving an evaluation function from a simple mechanical model,
called Johnson-Segalman (JS) equation with a diffusion term (DJS equation).  The DJS
model is a widely accepted model as a paradigm of shear banding, with which
characteristic experimental results are well reproduced
numerically \cite{kn:yuan-Europhys,kn:Lu,kn:Fielding}. Here, the term ''mechanical''
indicates that the model is constructed in a purely mechanical manner, i.e., the model
does not have an explicit thermodynamic potential. While there are some other
extended models on shear bands \cite{doi-onuki,kn:coupling-model-yuan,
kn:Fielding-Olmsted-2003}, as a first trial on this issue we will adopt the simplest
DJS equation. In fact, the method we use in the present study, which is based on the
reduction theory developed in the field of non-linear
dynamics \cite{kn:kuramoto-book}, is also applicable to a wide class of models.

We start with the following governing equations for a viscoelastic fluid:
\begin{eqnarray}
\nabla \cdot \boldsymbol{v} \!\!\!&=&\!\!\! 0,
\label{12:rwoene-incompressible-condition}
\\
\rho (\frac{\partial }{\partial t}+\boldsymbol{v} \cdot \nabla) \boldsymbol{v}
\!\!\!&=&\!\!\!
\nabla \cdot (\boldsymbol{\Sigma} + 2 \eta \boldsymbol{D} - p \boldsymbol{I} ),
\label{13:rwoemR-momentum-balance-equation}
\\
(\frac{\partial }{\partial t}+\boldsymbol{v} \cdot \nabla) \boldsymbol{\Sigma }
\!\!\!&=&\!\!\!
a (\boldsymbol{D} \cdot \boldsymbol{\Sigma}+ \boldsymbol{\Sigma} \cdot
\boldsymbol{D})+( \boldsymbol{\Omega} \cdot \boldsymbol{\Sigma} -
\boldsymbol{\Sigma} \cdot \boldsymbol{\Omega} )+ 2 G \boldsymbol{D} - \frac{
\boldsymbol{\Sigma}}{\tau}- 2 D_0 \nabla^2 \boldsymbol{D}.
\label{11:rwoesi-DJS-equation}
\end{eqnarray}

Equation (\ref{12:rwoene-incompressible-condition}) represents the incompressibility
condition of the fluid, where $\boldsymbol{v}$ is the fluid velocity. This condition
holds in usual fluids and guarantees that the density $\rho$ is constant throughout
the fluid. Equation (\ref{13:rwoemR-momentum-balance-equation}) represents the
momentum balance of the fluid, where the total stress tensor denoted by
$\boldsymbol\sigma$ is assumed to be composed of two contributions as
$\boldsymbol{\sigma}=\boldsymbol{\Sigma}+2 \eta \boldsymbol{D},$ where
$\boldsymbol{\Sigma}$ is the contribution from the polymeric components and $2 \eta
\boldsymbol{D}$ is the contribution from the solvent.  We assume that the polymeric
component $\boldsymbol{\Sigma}$ obeys the DJS equation given by
Eq. (\ref{11:rwoesi-DJS-equation}), while the solvent is assumed to be a Newtonian
fluid so that its contribution is a product of the viscosity $\eta$ and the symmetric
part of the velocity gradient tensor $\boldsymbol{D}= ((\nabla \boldsymbol{v})^{T}+
\nabla \boldsymbol{v})/2$. {The quantity $p$ is the pressure, and $\boldsymbol{I}$
is the unit tensor.} Equation (\ref{11:rwoesi-DJS-equation}) is the DJS equation,
where $a$ is a parameter that satisfies $0 \le a \le 1$.  This parameter $a$ is
called the ''slip parameter'', which represents the degree of non-affine deformation
under the flow \cite{kn:original-JS-equation}. $\boldsymbol{\Omega}$ is the
antisymmetric part of the velocity gradient tensor, defined by $\boldsymbol{\Omega}=
((\nabla \boldsymbol{v})^{T} - \nabla \boldsymbol{v})/2$. The coefficients $G$,
$\tau$ and $D_0$ in Eq. (\ref{11:rwoesi-DJS-equation}) are scalar constants having
the dimensions of stress, time, and (stress $\times$ (length)${}^2$),
respectively. The last term $\nabla^2 \boldsymbol{D}$ is the diffusive stress term,
the components of which are $\sum_{i} {\partial}^2D_{jk}/\partial x_i^2$ for any $j$ and
$k$.  {In References \cite{kn:Lu,kn:Fielding}, the non-local term of the DJS
equation is assumed to have the form $\nabla^2 \boldsymbol{\Sigma}$.  The physical
meaning of this form can be understood as a diffusion of the stress tensor
originating from the diffusive flux of the polymeric components.  Although the
physical meaning of $\nabla^2 \boldsymbol{D}$ appears to be less obvious, it
will be shown that the final results of the present analysis are unchanged both of these
choices.} (See the last paragraph of this article.)

Using this setup, we consider a one-dimensional simple shear flow, i.e., a flow
contained between two parallel plates moving with a constant relative
speed $v_0$ along the $x$-axis. The velocity distribution of the fluid
$\boldsymbol{v} \equiv (v_x,v_y,v_z)$ in such a situation is given as $v_x=v_x(y,t)$
$v_y=v_z=0$ with the boundary conditions $v_x(0,t)=0$ and $v_x(L,t)=v_0$ for any time
$t$, where $L$ is the distance between the two plates. For the solvability of the
model, we need additional boundary conditions. We require ${\partial}^2
v_x(y,t)/\partial y^2|_{y=0,L}=0$ for any time $t$. This boundary condition
corresponds to the situation in which the polymeric components are impenetrable at the
walls of the plates \cite{kn:meaning-of-diffusion-term,kn:meaning-of-diffusion-term-olmsted}.

To simplify the equations, we use the same non-dimensionalization procedure
introduced in early investigations \cite{kn:nondimensionalization,kn:Lu},
i.e., $t/\tau \rightarrow t$, $y/L \rightarrow y$, $\rho L^2/\tau^2 G \rightarrow \rho
$, $\eta/G \tau \rightarrow \eta $ and $D_0/2 G L^2 \rightarrow D_1$.  By applying
this non-dimensionalization, we obtain three non-dimensional quantities,
$\kappa(y,t)$, $N(y,t)$ and $S(y,t)$, defined by $\kappa \equiv \sqrt{1-a^2} \tau
\frac{\partial v_x}{\partial y}$, ${N}\equiv [(1-a)\Sigma_{xx} - (1+a)
\Sigma_{yy}]/2G$, and ${S} \equiv\sqrt{1-a^2} \Sigma_{xy}/G$.  Then,
Eqs. (\ref{12:rwoene-incompressible-condition})-(\ref{11:rwoesi-DJS-equation}) reduce
to the following set of equations for $\kappa$, $N$ and $S$: \begin{eqnarray}
\frac{\partial {\kappa}}{\partial {t}} \!\!\!&=&\!\!\! \frac{1}{{\rho}}
\frac{\partial {}^2}{\partial {y}^2} ({S}+ {\eta} {\kappa}),
\label{5:rwuppx-time-evolution-of-kappa} \\ \frac{\partial {N}}{\partial {t}}
\!\!\!&=&\!\!\! {\kappa} {S}- {N}, \label{6:rwupof-time-evolution-of-N} \\
\frac{\partial {S}}{\partial {t}} \!\!\!&=&\!\!\! -{\kappa} {N}+ {\kappa}- {S }- D_1
\frac{\partial {}^2}{\partial {y}^2} {\kappa}. \label{8:rwupnp-time-evolution-of-S}
\end{eqnarray}. Here, Eq. (\ref{5:rwuppx-time-evolution-of-kappa}) is obtained by
differentiating the $x$-component of both sides of
Eq. (\ref{13:rwoemR-momentum-balance-equation}) and eliminating $v_x$.  In this
non-dimensional system, the boundary conditions are given as \begin{equation}
\int_{0}^{1} {\kappa}({y}, t) d {y}=\bar{\kappa} \qquad \mbox{and} \qquad
\left. \frac{\partial \kappa({y},t)}{\partial {y}} \right|_{{y}=0,1}=0,
\label{20:rwodjO-boundary-condition-smooth} \end{equation} where $\bar\kappa$ is a
constant given by $\bar{\kappa}=\tau \sqrt{1-a^2} v_0/L$.  In the following analysis,
we assume that the quantity $D_1$ is an extremely small positive number compared to
unity, which may be justified by the fact that $D_1 \propto 1/L^2$. This assumption
corresponds to the situation in which the width of the interface between the shear bands
is much smaller than the system size because in this model the width of the interface
is given by a quantity proportional to $\sqrt{D_1}$.

A trivial $y$-independent stationary solution of
Eqs. (\ref{5:rwuppx-time-evolution-of-kappa})-(\ref{8:rwupnp-time-evolution-of-S})
that satisfies the boundary conditions
of Eqs. (\ref{20:rwodjO-boundary-condition-smooth}) is $\kappa=\bar{\kappa}$,
$N=\bar{\kappa}^2/(1+\bar{\kappa}^2)$, and $S=\bar{\kappa}/(1+\bar{\kappa}^2)$.  This
solution gives a trivial value of the total stationary shear stress $\bar{\sigma}=
\frac{\bar{\kappa}}{(1+\bar{\kappa}^2)} + \eta \bar{\kappa}$. When $\eta > 1/8$,
$\bar{\sigma}$ is a monotonic function of $\bar{\kappa}$, and when $\eta < 1/8$, $\bar{\sigma}$ is a non-monotonic N-shaped function of $\bar{\kappa}$ that has a region with a
negative slope. The negative-slope region of the DJS equation is given by
$\kappa_{n-} < \bar\kappa< \kappa_{n+}$ with $\kappa_{n\pm}=\sqrt{-1+\frac{1}{2 \eta}
\pm \frac{\sqrt{1-8\eta}}{2 \eta}}$. The trivial stationary
solutions have been proven to be unstable in this region \cite{kn:unstable,kn:unstable-yuan}. When $\eta$ approaches $1/8 $, the negative-slope region $(\kappa_{n-},\kappa_{n+})$ converges to
a single point $\bar\kappa=\sqrt{3}$, which allows us to define the ``critical
point'' of the DJS equation as $({\bar\kappa}^*,\eta^*)=(\sqrt{3},1/8)$. At this
critical point, the conditions $d\bar\sigma / d\bar\kappa|_{\bar\kappa={
\bar\kappa}^*}=0$ and $d^2\bar\sigma/d\bar\kappa^2|_{\bar\kappa={\bar\kappa}^*}=0$
hold simultaneously.

A non-trivial stationary solution of
Eqs.(\ref{5:rwuppx-time-evolution-of-kappa})-(\ref{8:rwupnp-time-evolution-of-S}) in
the negative-slope region for $\eta<1/8$ is obtained in the case with the $\nabla^2
\boldsymbol{D}$ term in Eq. (\ref{11:rwoesi-DJS-equation}). By assuming a stationary
solution of
Eqs.(\ref{5:rwuppx-time-evolution-of-kappa})-(\ref{8:rwupnp-time-evolution-of-S})
under the boundary condition $v(y,t)\vert_{y=0, L} = const$, we obtain
\begin{equation} \kappa (1-\kappa (\sigma_s-\eta \kappa))-(\sigma_s-\eta \kappa)-D_1
\frac{ d^2 \kappa}{d y^2}=0.  \label{21:rwodhQ-newton-equation-for-kappa}
\end{equation} We can identify this equation with an equation of motion for a
particle with position $\kappa$ at time $y$ moving in a quartic potential.  This
equation has a non-trivial solution that satisfies the boundary conditions Eq.(\ref
{20:rwodjO-boundary-condition-smooth}) only when \begin{equation} \sigma_s=3
\sqrt{-\eta^2+ \frac{\eta}{2}}.  \label{9:rwoewy-exact-sigma-select} \end{equation}
Note that this value of $\sigma_s$ does not depend on $\bar{\kappa}$. At this value
of $\sigma_s$, the non-trivial solution of
Eq. (\ref{21:rwodhQ-newton-equation-for-kappa}) is given as \begin{equation}
\kappa(y)=\kappa_0(y) \equiv \frac{\kappa_+-\kappa_-}{2} \tanh \left[ \frac{ y-
\displaystyle \frac{\kappa_+-\bar{\kappa}}{\kappa_+-\kappa_-}}{ \displaystyle \xi}
\right] +\frac{\kappa_+ + \kappa_-}{2}, \label{14:ruyrHJ-form-of-kappa}
\end{equation}, where $\kappa_{\pm}=\frac{\sqrt{\frac{1}{\eta}-2} \pm
\sqrt{\frac{1}{\eta}-8}}{\sqrt{2}}$. This non-uniform solution describes a shear
banding, in which the relative volume fraction of the regions with lower shear rate
is given by $(\kappa_+-\bar{\kappa})/(\kappa_+-\kappa_-)$ and the width of the
interface between different regions is given by $\xi \equiv \sqrt{D_1/(1-8
\eta)}$. From the symmetry of Eq.(\ref{21:rwodhQ-newton-equation-for-kappa}), we know
that $\kappa(y)=\kappa_0(1-y)$ is also a non-uniform solution of
Eq. (\ref{21:rwodhQ-newton-equation-for-kappa}). It is worth noting that the 
analytical results of Eqs.(\ref{9:rwoewy-exact-sigma-select}) and
(\ref{14:ruyrHJ-form-of-kappa}) are in good agreement with the numerical results for
the same model \cite{kn:yuan-Europhys}.

As mentioned in the introduction, a linear stability analysis shows that the trivial
homogeneous solution is unstable in the negative-slope region ($\kappa_{n-}< \kappa<
\kappa_{n+}$), and therefore a nontrivial solution should emerge.  However, in the
remaining regions ($\kappa_- <\kappa< \kappa_{n-}$ and $ \kappa_{n+}
<\kappa<\kappa_+$), where $\bar\sigma$ has a positive slope, there is no guarantee
that the non-trivial solution is finally realized because the homogeneous flow is
still stable (at least locally) in these regions. Therefore, for a complete understanding
of the stress plateau, it is not sufficient to determine the value of $\sigma_s$ uniquely
and obtain the corresponding non-trivial solution. The metastability of the homogeneous flow in the positive-slope regions must be demonstrated. To do this, we shall reduce the DJS equation in the neighborhood of its critical point.

The reduction procedure is as follows. First, we ignore the inertia term (the adiabatic approximation) in Eq. (\ref{5:rwuppx-time-evolution-of-kappa}) because the Reynolds number is small for usual viscoelastic fluid flows. We then have an approximate expression for $\kappa$ as $ \kappa(y,t) \simeq \bar{\kappa}-e (S-S_0), $ where $e=1/\eta$, and $S_0$ is the mean value of $S$ over the spatial coordinate $y$, i.e., $S_0(t)=\int_{0}^{1} S(y,t) dy$. Substituting this approximate
expression for $\kappa$ into Eqs.(\ref{6:rwupof-time-evolution-of-N}) and
(\ref{8:rwupnp-time-evolution-of-S}) yields \begin{eqnarray} \frac{\partial
N}{\partial t} \!\!\!&=&\!\!\! (\bar{\kappa}-e (S-S_0))S-N,
\label{23:rwodYq-time-evolution-equation-for-N-after-adiabatic-approximation} \\
\frac{\partial S}{\partial t} \!\!\!&=&\!\!\! (\bar{\kappa}-e (S-S_0))(1-N)-S+d
\frac{\partial {}^2S}{\partial y^2},
\label{24:rwodXs-time-evolution-equation-for-S-after-adiabatic-approximation}
\end{eqnarray} where $d=D_1 e$. Next, in order to clarify which mode is dominant in
the critical region, we carry out a linear stability analysis at the critical point
$(\bar{\kappa}^{*},e^{*})=(\sqrt{3},8)$. Linearizing
Eqs.(\ref{23:rwodYq-time-evolution-equation-for-N-after-adiabatic-approximation}) and
(\ref{24:rwodXs-time-evolution-equation-for-S-after-adiabatic-approximation}) with respect to
the stationary solution at the critical point,
$({\bar{N}}^*,{\bar{S}}^*)=(3/4,\sqrt{3}/4)$, in terms of the deviations $s_0$,
$n_0$, $\hat{s}$, and $\hat{n}$ defined by $s_0=\int_{0}^{1} (S-{\bar{S}}^*) dy$,
$n_0=\int_{0}^{1} (N-{\bar{N}}^*) dy$, $\hat{s}=S-{\bar{S}}^*-s_0$, and
$\hat{n}=N-{\bar{N}}^*-n_0$, we have \begin{equation} \left\{ \begin{split}
\frac{\partial n_0}{\partial t} &= -n_0+\sqrt{3} s_0 \\ \frac{\partial s_0}{\partial
t} &= -\sqrt{3} n_0 -s_0 \end{split} \right.  \hskip 3cm \left\{ \begin{split}
\frac{\partial \hat{n}}{\partial t} &= -\hat{n}-\sqrt{3} \hat{s} \\ \frac{\partial
\hat{s}}{\partial t} &= -\sqrt{3} \hat{n} - 3 \hat{s}.  \end{split} \right.
\label{17:rughld-later-set-of-equations} \end{equation} Here, we have neglected the
diffusion term because, according to the reduction theory for partial differential
equations, the diffusion term should be regarded as a small perturbation to the
uniform system \cite{kn:kuramoto-book}.  The eigenvalues of the former set of
equations are $-1 \pm i \sqrt{3}$, and those for the latter set of equations are $0$
and $-4$. The corresponding eigenmodes of the latter set of equations are
\begin{eqnarray} \phi \!\!\!&=&\!\!\! \frac{1}{2}( - \sqrt{3} \hat{n} + \hat{s}),
\label{16:ruyhH4-definition-of-phi} \\ \psi \!\!\!&=&\!\!\! \frac{1}{2}( \hat{n} +
\sqrt{3} \hat{s}), \label{30:rugheu-psi-definition} \end{eqnarray} respectively.
Thus, we find that, in the vicinity of the critical point, the variable $\phi$ is the
unique slow one  of the system and the other variables $n_0$, $s_0$ and $\psi$ are
solved by $\phi$ \cite{kn:kuramoto-book,kn:Haken-book}. Based on this observation, we
first rewrite the original
Eqs.(\ref{23:rwodYq-time-evolution-equation-for-N-after-adiabatic-approximation}) and
(\ref{24:rwodXs-time-evolution-equation-for-S-after-adiabatic-approximation}) in
terms of $n_0$, $s_0$, $\phi$ and $\psi$ as \begin{eqnarray} \frac{\partial
n_0}{\partial t} \!\!\!&=&\!\!\! {\bar{S}}^* (\bar{\kappa}-{ \bar{\kappa}}^*) -n_0
+\bar{\kappa} s_0 - \frac{e}{4} \langle (\phi+\sqrt{3} \psi)^2 \rangle,
\label{25:rwodSb-n0-dot} \\ \frac{\partial s_0}{\partial t} \!\!\!&=&\!\!\!
(\bar{\kappa}-{\bar{\kappa}}^*) (1-{\bar{N}}^*) - \bar{\kappa} n_0- s_0 + \frac{e}{4}
\langle (-\sqrt{3} \phi+ \psi)(\phi+\sqrt{3} \psi) \rangle, \label{26:rwodSA-s0-dot}
\\ \frac{\partial \phi}{\partial t} \!\!\!&=&\!\!\! \frac{1}{2} (-\sqrt{3} f_1 +f_2)
+\frac{d}{4} \partial_y^2 (\phi+\sqrt{3} \psi), \label{27:rwodRt-phi-dot} \\
\frac{\partial \psi}{\partial t} \!\!\!&=&\!\!\! \frac{1}{2} (f_1 +\sqrt{3} f_2)+
\frac{\sqrt{3} d }{4} \partial_y^2 (\phi+\sqrt{3} \psi) , \label{28:rwodRd-psi-dot}
\end{eqnarray} where the bracket denotes an integral over the entire domain of $y$,
i.e., $\langle \cdot \rangle=\int_{0}^{1} \cdot dy$.  The quantities $f_1$ and $f_2$
are functionals of $n_0$, $s_0$, $\phi$, and $\psi$, defined by $f_1=\{-\hat{n}+(-e
{\bar{S}}^* + \bar\kappa) \hat{s} -e s_0 \hat{s} -e (\hat{s}^2-\langle \hat{s}^2
\rangle)\}|_p$ and $f_2=\{-\bar{\kappa} \hat{n}-(e (1-{\bar{N}}^*) +1 ) \hat{s} +e
n_0 \hat{s} +e (\hat{n} \hat{s}-\langle \hat{n} \hat{s} \rangle)\}|_p$, where the
subscript $p$ indicates the substitutions $\hat{n}=\frac{1}{2}( - \sqrt{3} \phi +
\psi) $ and $\hat{s}=\frac{1}{2}( \phi + \sqrt{3} \psi)$ in each expression.  Note
that these Eqs.(\ref{25:rwodSb-n0-dot})-(\ref{28:rwodRd-psi-dot}) are still
equivalent to the original
Eqs.(\ref{23:rwodYq-time-evolution-equation-for-N-after-adiabatic-approximation}) and
(\ref{24:rwodXs-time-evolution-equation-for-S-after-adiabatic-approximation}). Next,
we assume that the variables $n_0$, $s_0$, and $\psi$ are functionals of $\phi$ and
functions of $y$, denoted by $n_0[\phi,y]$, $s_0[\phi,y]$ and $\psi[\phi,y]$, and
define a functional $G[\phi,y]$ as the right-hand side of
Eq.(\ref{27:rwodRt-phi-dot}). Substituting these forms into
Eqs. (\ref{25:rwodSb-n0-dot})-(\ref{28:rwodRd-psi-dot}) gives four algebraic
equations for $n_0$, $s_0$, $\psi$, and $G$ with independent variables $\phi$ and
$y$. Here, our task is reduced to determining the forms of these four
functionals. Introducing two small parameters $k$ and $\epsilon$ defined as
$k=\bar{\kappa}-{\bar{\kappa}}^*$ and $\epsilon=e-e^*$, ($|\epsilon| \ll 1$, $|k| \ll
1$) and realizing that the difference between the shear rates of each domain is on
the order $O(\sqrt{\epsilon})$ $(\epsilon >0 )$, which is known from the expanded
form of $\bar\sigma$ near the critical point, we can proceed to the perturbative
calculation systematically. The results are $n_0=\frac{\sqrt{3}}{8} k
-\frac{k^2}{8}-2 \langle \phi^2 \rangle $, $ s_0=-\frac{k}{8} $, $\psi=
\frac{k\phi}{2}-(\phi^2-\langle \phi^2 \rangle) $ and
$G=(\frac{\epsilon}{8}-\frac{3}{4} k^2) \phi +3 k (\phi^2-\langle \phi^2 \rangle)-4
(\phi^3-\langle \phi^3 \rangle)$. Here we have used the property $\langle \phi
\rangle=0$, which results from its definition (\ref{16:ruyhH4-definition-of-phi}). In
the above calculation, we have neglected the diffusion terms because of the smallness
of $d$. If we retain the diffusion terms up to the lowest order in $d$, we have the
following time evolution equation for $\phi$: \begin{equation} \frac{\partial
\phi}{\partial t}=(\frac{\epsilon}{8}-\frac{3}{4} k^2) \phi +3 k (\phi^2 - \langle
\phi^2 \rangle)-4 (\phi^3 - \langle \phi^3 \rangle) + \frac{d}{4} \frac{\partial
{}^2\phi}{\partial y^2}.  \label{10:rwoetw-main-result} \end{equation} This equation
expresses the main result of the present research. The dynamic variable $\phi$ is
essentially an ``order parameter'' that which measures the inhomogeneity of the
viscoelastic stress field and includes the contributions from both the shear stress
and the first normal stress difference.  Equation (\ref{10:rwoetw-main-result}) has the same form as that of the time-dependent Ginzburg-Landau (TDGL) equation for first-order phase transitions in
thermodynamics \cite{kn:landau-book,kn:book-ohta}. A non-trivial stationary solution of
Eq. (\ref{10:rwoetw-main-result}) for $\epsilon>0$ is readily obtained, which is
consistent with the exact solution of Eq. (\ref{14:ruyrHJ-form-of-kappa}) to the lowest
order in $\epsilon$. An important advantage of the reduced equation, Eq.
(\ref{10:rwoetw-main-result}) is that we are able to prove the metastability of the
homogeneous flow, which is readily performed by conventional means, e.g., a common
tangent construction. A detailed discussion on the properties of
Eq. (\ref{10:rwoetw-main-result}) will be reported in a future paper \cite{kn:in-preparation}.

Finally, we would like to comment on our choice of the diffusion term. In
Eq. (\ref{11:rwoesi-DJS-equation}), even if we replace the diffusion term $D_0
\nabla^2 \boldsymbol{D}$ with the usual form $D_2 \nabla^2 \boldsymbol{\Sigma}$, the
reduced TDGL-type equations is unchanged. The only difference is the diffusion
constant appearing in (\ref{10:rwoetw-main-result}), where $d/4$ is replaced by $\tau
D_2/L^2$.

\acknowledgements The authors would like to thank T. Shibata, Y. Kuramoto,
T.Yamaguchi, T. Chawanya, N. Uchida, T. Ohta, H. Watanabe, and T. Koga for their helpful discussions
and comments. The present study was supported by Grant-in-Aid for Scientific Research on
Priority Area ``Soft Matter Physics'' from the Ministry of Education, Culture,
Sports, Science, and Technology of Japan.

\newpage \begin{figure}[ht] \centering
\includegraphics[width=10cm,clip]{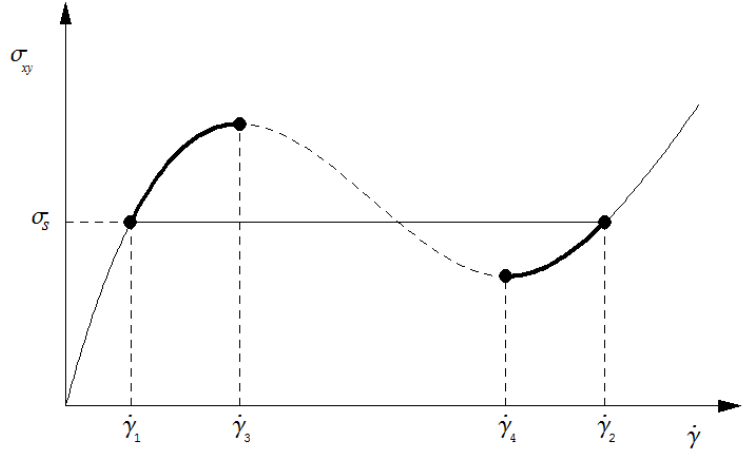}
\caption{ Figure 1: Schematic representation of shear banding. In the regions
of shear rate where the SS curve has a positive slope, $\dot\gamma_1 <
\dot\gamma < \dot\gamma_3$, and $\dot\gamma_4 < \dot\gamma < \dot\gamma_2$, both
the homogeneous flow and banded flow are stable. Hence, in order to determine which
state is finally realized in these regions, it is necessary to have some
evaluation function to compare the stabilities of these states.} \end{figure}

\end{document}